\documentclass[12pt,titlepage]{article}
\usepackage{setspace} 
\usepackage{amsmath}
\usepackage{bm}
\usepackage{pslatex}
\usepackage{times}
\usepackage{graphicx}
\usepackage{subfig}

\usepackage[round,numbers,sort&compress]{natbib} 
\title{Dissipative electro-elastic network model of protein electrostatics}

\author{Daniel R. Martin, S.\ Banu Ozkan, and 
Dmitry V.\ Matyushov\footnote{Correspondence: dmitrym@asu.edu} \\
Center for Biological Physics, Arizona State University, \\ 
PO Box 871504, Tempe, AZ 85287-1504
}

\date{\today}
\begin{document}

\maketitle

\abstract{ABSTRACT \ We propose a dissipative electro-elastic network
  model (DENM) to describe the dynamics and statistics of
  electrostatic fluctuations at active sites of proteins. The model
  combines the harmonic network of residue beads with overdamped
  dynamics of the normal modes of the network characterized by two
  friction coefficients. The electrostatic component is introduced to
  the model through atomic charges of the protein force field. The
  overall effect of the electrostatic fluctuations of the network is
  recorded through the frequency-dependent response functions of the
  electrostatic potential and electric field at the active site. We
  also consider the dynamics of displacements of individual residues
  in the network and the dynamics of distances between pairs of
  residues. The model is tested against loss spectra of residue
  displacements and the electrostatic potential and electric field at
  the heme's iron from all-atom molecular dynamics simulations of
  three hydrated globular proteins.

 }

\clearpage

\section*{INTRODUCTION}
Proteins in solution exist not as static structures but instead as
dynamic entities interconverting between conformational sub-states on
the time-scale reflecting the activation barriers involved in the
transitions \citep{Fenimore:04,HenzlelWildman:07,Frauenfelder:09}. It
is becoming increasingly clear that biomolecules are fluctuating
machines, using dynamical equilibria between their conformational states
to promote their function \cite{HenzlelWildman:07,Marlow:2010bh}.  The
flexibility of the protein structure naturally leads to changes in
both the charge distribution of the protein itself and in the
polarization of the interfacial waters following the protein
motions. These dynamically fluctuating polarization fields will affect
many properties sensitive to electrostatics, including the rates of
enzymatic reactions \cite{Warshel:06}. The dynamic nature of proteins,
enhanced by the conformational manifold provided by hydration water
\citep{Fenimore:04}, requires the shift of the focus from statistical
averages and corresponding thermodynamic parameters, such as
equilibrium Gibbs energies, to entire statistical distributions of
observables and their dynamics expressed through time correlation
functions.

This requirement is certainly the case for the problem of protein
electron transfer, where, according to the standard Marcus picture
\citep{MarcusSutin}, both the average of the energy gap between the
donor and acceptor energy levels of the electron and its fluctuation
are required to determine the probability of electron tunneling. This
is just one example when the knowledge of the entire fluctuation
spectrum of the observables is of significant interest
\citep{DMpccp:10}.  Other applications, such as electrostatic
fluctuations reported by broad-band dielectric
spectroscopy \cite{Khodadadi:2010qf}, THz absorption
\cite{Ebbinghaus:07}, and light scattering techniques
\cite{Perticaroli:10} require either the entire fluctuation spectrum
or a set of cumulants of the corresponding experimental observables.

With the general focus on the fluctuation spectrum of proteins, we
propose here a model for the calculation of the statistics and
dynamics of the electrostatic potential and electric field in
proteins.  Electrostatic interactions are clearly important for
protein stability and function \cite{Warshel:06}. Electrostatic
solvation and interactions affect the stability of folded proteins and
their aggregation and crystallization
\citep{Richardson:02,Lawrence:07,Pace:09}. Electrostatics might also
play a significant role in promoting high-temperature flexibility of
proteins through the coupling of the charge distribution in the
protein to the electrostatic fluctuations produced by the
protein-water interface \citep{DMpre:11}.

The dynamics of electrostatic fluctuations spans an enormous range of
time-scales from sub-picoseconds to sub-microseconds, and possibly
longer \cite{Berg:05,Tripathy:10,Khodadadi:2010qf}. While
spectroscopic techniques are capable of recording the ultra-fast
fs-to-ps relaxation \cite{Zhang:07}, the fluctuations of the
protein-water interface recorded by M{\"o}ssbauer spectroscopy and
neutron scattering cover much longer time-scales from
sub-nanoseconds to sub-microseconds \citep{Frauenfelder:09}. Our
present model aims at these long, and perhaps even longer, time-scales
by coarse-graining the protein into an elastic network of beads, each
representing a protein residue.

Elastic network models have consistently shown good performance in
characterizing global, large-scale motions of proteins
\citep{Tirion:96,Atilgan:01,Tama:2001tg,Tama:2003hc,Tama:2005oq,Lu:2006cr,Moritsugu:2007ff,Lu:2008dq,Riccardi:09,Bahar:10aa}. While
the low-frequency portion of the spectrum of protein motions is mostly
determined by the distribution of mass and molecular shape
\citep{Halle:02,Lu:2005mi,Tama:06}, improvements are still needed to
account for deficiencies of networks when localized protein motions
are involved \citep{Petrone:06,Lyman:08,hinsen:10766,Miller:2008uq}.
Electrostatic interactions, on the other hand, are notoriously
long-ranged. The electrostatic potential, slowly changing over a
nanometer-size biomolecule, effectively averages out local structural
variations and is mostly sensitive to the global distribution of
charge within the molecule. From this perspective, the combination of
even a basic elastic network with the molecular charge distribution
might be sufficient to describe the statistics of the potential
fluctuations and their slow dynamics.

We propose here a model combining the distribution of molecular charge
from the standard atomic force-fields with the dynamics and statistics
of protein motions derived from an elastic network.  To test the
model, we employ all-atom Molecular Dynamics (MD) simulations of three
hydrated heme proteins. Previous analysis of these data has emphasized
the importance of nanosecond (ns) relaxation modes in the
electrostatic fluctuations of the protein matrix and the protein-water
interface \citep{DMpre:11,DMjpcb:11}. This time-scale is relevant to
biological function since heme proteins are typical components of
biology's energy chains transporting electrons on
nanosecond-to-microsecond time-scales \cite{DMjpcb:09,DMpccp:10}. The
nanosecond time-scale fluctuations are also consistent with the
dynamics of the global motions of the network of residues. An elastic
network model naturally fits the physics of the problem. The present
contribution is a first step in developing network models of the
electrostatic response of hydrated proteins. Here, we focus on the
electrostatics of the protein matrix only, leaving the water component
of the overall response to future studies.

\section*{THEORY AND METHODS}
\small
\subsection*{Elastic network model}
A coarse-grained elastic network model (ENM) assigns a node to a
collection of atoms reducing the computational burden of an all-atom
normal-mode analysis. The typical coarse-graining is done on the level
of individual aminoacids (1-bead model \citep{Tozzini:11}). The
position of the node is defined by the coordinates of the C$^{\alpha}$
atom. The springs connecting the nodes represent the bonded and
non-bonded interactions within an accepted cutoff distance
\cite{Tirion:96,Atilgan:01,Tama:2001tg} or by a distance-dependent force
constant
\citep{hinsen:10766,Hinsen:2008bh,Riccardi:09,Gerek:2011ly}. The ENM
diagonalizes the Hessian matrix $\mathbf{H}$ derived from a simplified
Hookean potential suggested by Tirion \cite{Tirion:96}. This
potential, $E_{ij} = C(r_{ij} -r_{0,ij})^2/2$, describes the
elongation $r_{ij}=|\mathbf{r}_i-\mathbf{r}_j|$ between nodes $i$ and
$j$ in the network characterized by one universal force constant $C$
and the structural information stored in the equilibrium bead
positions $\mathbf{r}_{0,i}$
($r_{0,ij}=|\mathbf{r}_{0,i}-\mathbf{r}_{0,j}|$).

The Hessian $H_{ij}^{\alpha\beta}$ is a $3N\times 3N$ matrix
representing the protein elastic energy as a quadratic form of the
Cartesian displacements $\delta r_{i}^{\alpha}=r_i^{\alpha}-
r_{0,i}^{\alpha}$ of individual beads in the network
\begin{equation}
  \label{eq:2}
    E = (C/2) \sum_{i,j} H_{ij}^{\alpha\beta} \delta
    r_{i}^{\alpha}\delta r_{j}^{\beta}  .
\end{equation}
Here, $i$ and $j$ run between 1 and $N$, $\alpha,\beta$ indicate the
Cartesian projections, and summation over repeated Greek indices is
assumed here and throughout. The Hessian is diagonalized by the
unitary matrix $\mathbf{U}$ producing $3N$ eigenvalues
$\lambda_m$. This standard linear algebra formalism
\citep{Tirion:96,Atilgan:01} yields the statistical correlator between
the displacements of residues $i$ and $j$ in the network
\begin{equation}
  \label{eq:1}
  \langle \delta r_i^{\alpha} \delta r_j^{\epsilon} \rangle = (\beta
  C)^{-1} \sum_{m,\gamma} U_{mi}^{\gamma\alpha} \lambda_m^{-1}
  U_{mj}^{\gamma\epsilon} ,  
\end{equation}
where $\beta=1/(k_{\text{B}}T)$ is the inverse temperature.

If the network is characterized by a universal force constant and a
cutoff, it yields a bell-shaped density of vibrational states. A
refinement of that network by adopting a stronger coupling between
covalently bound neighbors splits the density of states into two
maxima, in better agreement with all-atom calculations
\cite{Ming:2005fj,Moritsugu:2007ff,Gerek:2009ve}. This approximation
is adopted in our present calculations: the spring constant was
multiplied by a constant factor $\varepsilon=100$ for covalent
neighbors. In addition, a uniform cut-off radius of 15 \AA\ was used
in all calculations.

\subsection*{Overdamped network dynamics}
\label{2-2}
The standard mechanical ENM outlined above obviously lacks dissipative
dynamics. The equations of motion are harmonic, implying oscillatory
time correlation functions. In contrast, most correlation functions of
hydrated proteins observed by scattering
\citep{Smith:91,Khodadadi:2010qf,Perticaroli:10} and relaxation
\citep{Cametti:2011ys} techniques are exponential, corresponding to the
overdamped (Debye) dynamics, or stretched-exponential
\citep{Khodadadi:2010qf}. Several approaches can be implemented to
incorporate dissipative dynamics into the mechanical network of
beads. Langevin equations of motion for the ENM potential, within the
general framework of the Lamm-Szabo formalism \citep{Lamm:86}, have
been suggested \citep{Miller:2008uq,Essiz:2009fk}. This approach, and
some early suggestions \citep{ansari:1774,Erkip:04}, still requires
parameterization (done by fitting the rotational and translational
diffusion coefficients from MD) and, in addition, doubles the size of the matrix to
be diagonalized. We have adopted here a more straightforward formalism
not requiring additional computational resources.

Instead of a harmonically oscillating network, we have assumed for
each normal mode $\mathbf{q}_m$, diagonalizing the network's Hessian,
an overdamped motion described by the dissipative memory kernel
$\zeta(t)$ \citep{Hansen:03}. The equation of motion for such
overdamped dynamics is
\begin{equation}
  \label{eq:3}
  \int_0^{t} \zeta(t-t') \mathbf{\dot{q}}_m(t') dt' + \lambda_m
  \mathbf{q}_m = \mathbf{F}(t),
\end{equation}
where $\mathbf{F}(t)$ is an external force. 

The distinction between Eq.\ \eqref{eq:3} and the Langevin network
\citep{Miller:2008uq,Essiz:2009fk} is worth emphasizing here. In the
latter, uncoupled Langevin dissipative equations are first assigned to
each bead of the network. However, it was noted that dissipative
equations for the beads are likely to become coupled when the Langevin
dynamics of beads are consistently derived by integrating out the fast
degrees of freedom in the equations of motion
\cite{Soheilifard:2011pi}. Given that the assignment of uncoupled
Langevin equations to the individual beads is phenomenological from
the onset, one can introduce similar phenomenology at a different
level of the theory. Here, in the spirit of the standard normal-mode
analysis, we introduce uncoupled dissipative equations for the normal
modes (Eq.\ \eqref{eq:3}). This is still a phenomenological assumption
requiring further testing, but it reduces to the standard normal-mode
analysis in the static limit. 

When Laplace-Fourier transform \citep{Hansen:03} is applied to Eq.\
(\ref{eq:3}), the displacement response function for collective mode
$\mathbf{q}_m$ follows
\begin{equation}
  \label{eq:5}
  \chi_m(\omega) = \left[ i\omega\tilde \zeta(\omega) + \lambda_m\right]^{-1} ,
\end{equation}
where $\tilde \zeta(\omega)$ is the Laplace-Fourier transform of the
friction kernel $\zeta(t)$. Extending this procedure to all normal
modes of the network, one obtains the response function of bead
displacements
\begin{equation}
  \label{eq:6}
   \chi_{ij}^{\alpha\beta} (\omega) = C^{-1} \sum_{m}
   U_{mi}^{\gamma\alpha} 
   \left[\lambda_m +
    i \omega \zeta(\omega) \right]^{-1} U_{mj}^{\gamma\beta} . 
\end{equation}
In this equation, the response function $\bm{\chi}_{ij}(\omega)$,
which is a rank-2 tensor, represents the displacement of residue $i$
of the protein due to an oscillatory force with frequency $\omega$
applied to residue $j$ and propagated through the network to residue
$i$. This response function is the basis of the dissipative
electro-elastic network model (DENM) proposed here. At $\omega=0$,
Eq.\ \eqref{eq:6} returns the standard result of the ENM given by Eq.\
\eqref{eq:1}.

\begin{figure}
  \centering
  \includegraphics[width=6cm]{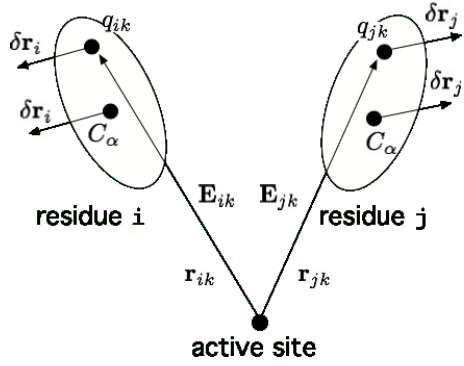}
  \caption{\small Cartoon illustrating the application of the
    dissipative electro-elastic network model (DENM) to the
    calculation of the electro-elastic response function at an atomic
    position in the active site of the protein. The elastic
    displacements of residues $i$ and $j$ create fluctuations of the
    electric fields by atomic charges $q_{ik}$ and $q_{jk}$.  Electric
    fields $\mathbf{E}_{ik}$ and $\mathbf{E}_{jk}$ are produced by the
    corresponding charges $q_{ik}$ and $q_{jk}$ at the active
    site. The overall electric field of residue $i$ at the active
    site, $\mathbf{E}_{0i}$, is obtained by summation of all
    $\mathbf{E}_{ik}$ produced by the charges $q_{ik}$ of the
    residue. }
  \label{fig:1}
\end{figure}

\subsection*{Electrostatic potential response function}
The ENM has shown good performance for low-frequency
structural/mechanical deformations of individual proteins and
biomolecular assemblies \citep{Bahar:10aa}. Our goal here is to
supplement these low-frequency motions with atomic partial charges to
model the dynamics and statistics of electrostatic fluctuations at
active sites of proteins.  Coarse-grained electrostatics of proteins
has been addressed in a number of recent papers
\citep{Skepo:2006gf,Pizzitutti:2007ul,Leherte:2009fr}, mostly dealing
with long-range protein assembly and interactions
\citep{Dong:08}. These approaches also involve coarse-graining of the
charge distribution \citep{Pizzitutti:2007ul,Leherte:2009fr} or
generating effective charges to reproduce the protein's external field
\citep{Berardi:2004pd}. Given that we are interested in the
electrostatic fluctuations, which are more sensitive to a local charge
distribution, atomic charges from force-field potentials are more
consistent with our purpose.

Our approach starts with solving the standard ENM problem to obtain a
set of displacements $\delta\mathbf{r}_{im}$ for each normal mode $m$
with the eigenvalue $\lambda_{m}$.  Each of these displacements is
then applied to all charges $q_{ik}$ of residue $i$, where index $k$
runs over all atoms of residue $i$ (Fig.\ \ref{fig:1}). Displacements
$\delta\mathbf{r}_{im}$ in turn produce dipole moments
$\delta\bm{\mu}_{ik}^{(m)} =q_{ik} \delta\mathbf{r}_{im}$ which
interact with charges of the protein active site. The overall
electrostatic effect of the fluctuating protein medium is given as a
sum over all contributions from each partial charge of all the
residues except for the active site. This cumulative effect of all
protein charges is given by the frequency-dependent response function
of the electrostatic potential $\chi_{\phi}(\omega)$
\begin{equation}
  \label{eq:7}
  \chi_{\phi}(\omega) = - \sum_{i,j}
  E_{0j}^{\alpha}\chi_{ij}^{\alpha\beta}(\omega) E_{0i}^{\beta} . 
\end{equation}
Here, $E_{0i}^{\alpha}$ is the $\alpha$ Cartesian projection of the
electrostatic field produced by all charges $q_{ik}$ of residue $i$ at an
atom in the active site (Fig.\ \ref{fig:1})
\begin{equation}
  \label{eq:8}
  E_{0i}^{\alpha} = -  \sum_k 
\frac{q_{ik} (\mathbf{r}_{ik} - \mathbf{r}_0)}{|\mathbf{r}_{ik} -
  \mathbf{r}_0|^3} .
\end{equation}
Here, $\mathbf{r}_0$ is the position of the atom in the active site
and $\mathbf{r}_{ik}$ is the position of charge $q_{ik}$ of residue
$i$.

\subsection*{Electrostatic field response function}
The potential response function $\chi_{\phi}(\omega)$ in Eq.\
(\ref{eq:7}) describes an oscillating electrostatic potential produced
by the protein matrix in response to placing an oscillatory charge at
the position in the active site where the potential is recorded (Fe
ion in our MD simulations). Similarly, the response function of the
electric field considers the field produced by the protein in response
to an oscillating dipole moment $\mathbf{m}_0(t) =
\mathbf{m}_0(\omega)\exp[i\omega t]$ placed in the active site.  The
deformation of the protein matrix induced by this probe dipole results
in the electric field $\mathbf{E}_0(\omega)$ at the same site. It can
be found by summation over the contributions from all individual
dipoles $\delta\bm{\mu}_{ik}(\omega)$ arising from residues' atomic
charges
\begin{equation}
  \label{eq:4}
  \mathbf{E}_0(\omega) = \sum_{ik} \mathbf{T}_{ik} \cdot  \delta\bm{\mu}_{ik} (\omega) .
\end{equation}
Here, $\mathbf{T}_{ik}$ is the dipolar tensor connecting the position
of the active site with the charge $q_{ik}$ of residue $i$. The dipole
moment $\delta\bm{\mu}_{ik}(\omega)$ at the position of $q_{ik}$ is
caused by the residue displacement $\delta \mathbf{r}_{ik}(\omega)$ and
can be expressed in terms of the displacement response function
\begin{equation}
  \label{eq:11}
  \delta \bm{\mu}_{ik}(\omega)=q_{ik}\sum_j \bm{\chi}_{ij}\cdot
  \delta \mathbf{F}_j(\omega), 
\end{equation}
where $\delta \mathbf{F}_j(\omega) = \sum_k q_{jk}\mathbf{T}_{jk}
\cdot \mathbf{m}_0(\omega)$ is the force caused by the dipole
$\mathbf{m}_0(\omega)$ at residue $j$. We therefore obtain a linear
relation between the dipole $\mathbf{m}_0(\omega)$ placed at the
active site and the electric field $\mathbf{E}_0(\omega)$ produced by
the protein in response to this perturbation. The proportionality
coefficient between the external perturbation and the response is the
response function given by the rank-2 tensor
\begin{equation}
  \label{eq:12}
  \chi_E^{\alpha\beta}(\omega)  = \sum_{i,j,k,l} q_{ik} T_{ik}^{\alpha\gamma}
  \chi_{ij}^{\gamma\delta}(\omega) T_{jl}^{\delta\beta} q_{jl} .
\end{equation}
Here, as above, the summation is done over the repeated Greek indices
denoting the Cartesian components of the corresponding tensors.  We
will be mostly interested in the trace of the tensor
\begin{equation}
  \label{eq:13}
  \chi_E(\omega) = \chi_E^{\alpha\alpha} (\omega) . 
\end{equation}

\subsection*{Response and correlation functions}
The response functions calculated by the DENM formalism need to be
related to time correlation functions supplied by the MD
trajectories. The dynamics of a general dynamical variable $\delta
X(t) = X(t) - \langle X \rangle$ is characterized by the time
self-correlation function
\begin{equation}
  \label{eq:15}
  C_X(t) = \langle \delta X(t)\delta X(0) \rangle .
\end{equation}
The normalized correlation function $S_X(t)=C_X(t)/ \langle(\delta X)^2\rangle$ is
related to the response function by the fluctuation-dissipation
equation which forms the basis of our analysis \cite{ChaikinLubensky}
\begin{equation}
  \label{eq:18}
  \chi_X(z) = \beta \langle(\delta X)^2\rangle \left[1 + i z\tilde S_X(z) \right],
\end{equation}
where $\tilde S_X(z)$ is the Laplace-Fourier transform of $S(t)$
defined in the upper half of the complex plane of $z$. 

Since our main focus is on variances of physical properties, we will
define the generalized compliance for the variable $X$ as follows
\begin{equation}
  \label{eq:19}
  \lambda_X = \beta\langle (\delta X)^2 \rangle /2 =
  \int_{0}^{\infty} \chi_X''(\omega) (d\omega/
  \pi\omega).    
\end{equation}
The same property can be calculated from the $\omega=0$ value of the
real part of the response function
\begin{equation}
  \label{eq:14}
  \lambda_X = (1/2) \chi_X'(0) .
\end{equation}
In case of residue displacements considered as variable $X$ ($X=r$),
$\lambda_r \propto C^{-1}$ is proportional to the inverse force
constant of the network springs. As for the compliance of a
macroscopic body, defined as the inverse of stiffness, $\lambda_r$
depends on the shape and boundary conditions, in contrast to elastic
moduli representing material properties.

For the elastic network, the generalized compliance reports on how the
softness of the network affects the variance of the variable of
interest.  In the case of $X=\phi$ reporting on the potential $\phi$
at the position of the heme's iron, $\lambda=e^2\lambda_{\phi}$ is the
reorganization energy of a half redox reaction corresponding to
changing the oxidation state of the heme
\cite{MarcusSutin,DMjcp2:08}. The standard definition of half-reaction
reorganization energy involves, instead of one atom, the distribution
of the electronic density of the transferred electron over a few atoms
of the active site. We simplify this problem here by assuming all
electron charge $e$ localized at the centroid of this charge density,
the iron atom of the heme. The electrostatic potential at a single
atom is a well defined physical property, but its variance is only an
approximate representation of the observable reorganization energy of
changing the redox state \cite{MarcusSutin}.

We want to gain insight into the dissipative dynamics of the elastic
network and its comparison with the dynamics calculated from MD
simulations.  The loss function $\chi_X''(\omega)$ provides direct
access to both the set of characteristic relaxation frequencies and
their relative contributions. However, it does not weigh the low and
high frequencies as they contribute to the variance in Eq.\
\eqref{eq:19}. Therefore, in addition to the loss function, we will
consider the function
\begin{equation}
  \label{eq:20}
  \alpha_X(\omega) = \frac{2}{\pi\omega}\
  \frac{\chi_X''(\omega)}{\chi_X'(0)} .
\end{equation}
This function is normalized, $\int_0^{\infty}\alpha_X(\omega) d\omega = 1$,
and tends to the characteristic relaxation time $\langle \tau \rangle$
in the $\omega\to 0$ limit
\begin{equation}
  \label{eq:21}
  \lim_{\omega\to 0} \alpha_X(\omega) = (2/ \pi) \langle \tau
  \rangle ,
\end{equation}
where
\begin{equation}
  \label{eq:22}
  \langle \tau \rangle = \int_0^{\infty} S_X(t) dt . 
\end{equation}

\subsection*{Proteins used in case studies and the simulation protocol}
Three heme proteins, oxidized form of myoglobin (metmyoglobin, metMB,
1YMB) reduced state of cytochrome B562 (cytB, 256B), and reduced state
of bovine heart cytochrome \textit{c} (cytC, 2B4Z) were simulated by
all-atom MD. The parameters of the proteins were taken from CHARMM27
force field \citep{charmm22}, and NAMD \citep{namd} was used for the
trajectories production. Each protein was placed at the center of a
cubic box with the side length of $\simeq 108$ \AA{} and solvated with
TIP3P waters \citep{tip3p:83}. The number of waters used in
simulations were: 32891 (metMB), 33268 (cytB), and 33189 (cytC).  The
VMD \cite{VMD} plugin Autoionize was used to neutralize the simulation
cell by adding Na$^+$ and Cl$^-$ ions to bring the ionic strength of
the solution to 0.1. Particle-mesh Ewald with the grid resolution $<1$
\AA{} was used for electrostatic interactions, and all other
non-bonded interactions were calculated within 12 \AA{}
cutoff. Following energy minimization, each protein was simulated for
5 ns in a NPT ensemble at $P=1$ atm and $T=300$ K. Temperature and
pressure were controlled by using the Langevin dynamics with the
damping coefficient of 5 ps$^{-1}$.  The production NVE trajectory was
started at the end of each 5 ns trajectory. The NVE ensemble is
required for the proper sampling of the long-time dynamics since we
found that using NPT and NVT ensembles artificially accelerates the
observables relaxing on sub-nanosecond to nanosecond time-scale
\citep{DMjpcb:10}. The integration step was 2 fs and simulation frames
for the analysis were saved each 0.05 ps. The simulation trajectories
were 65 ns (metMB), 100 ns (cytC), and 123 ns (cytB).

The force field standard parameters of the heme in the reduced state
are taken from CHARMM27 \citep{charmm22}. The charge of iron is
$q_{\text{Fe}}=0.24 $ in the reduced state. No standard parameters are
available for the oxidized heme required for the simulations of
metmyoglobin. These force field parameters were taken from Autenrieth
\textit{et al} \citep{Autenrieth:04}. The atomic charge of oxidized Fe
is $q_{\text{Fe}}=1.34$ in this parametrization. The iron atom is
five-coordinate in its oxidized state. In order to allow the iron to
shift out of the porphyrin plane, the harmonic force constant of the
bond between Fe and N$_{\epsilon}2$ of histidine 93 was adjusted to 65
kcal/mol as suggested in Ref.\ \cite{Meuwly:02} and also implemented
in Ref.\ \cite{ZhangStraub:09}.

\begin{figure}
  \centering
   \subfloat{\includegraphics*[width=5cm]{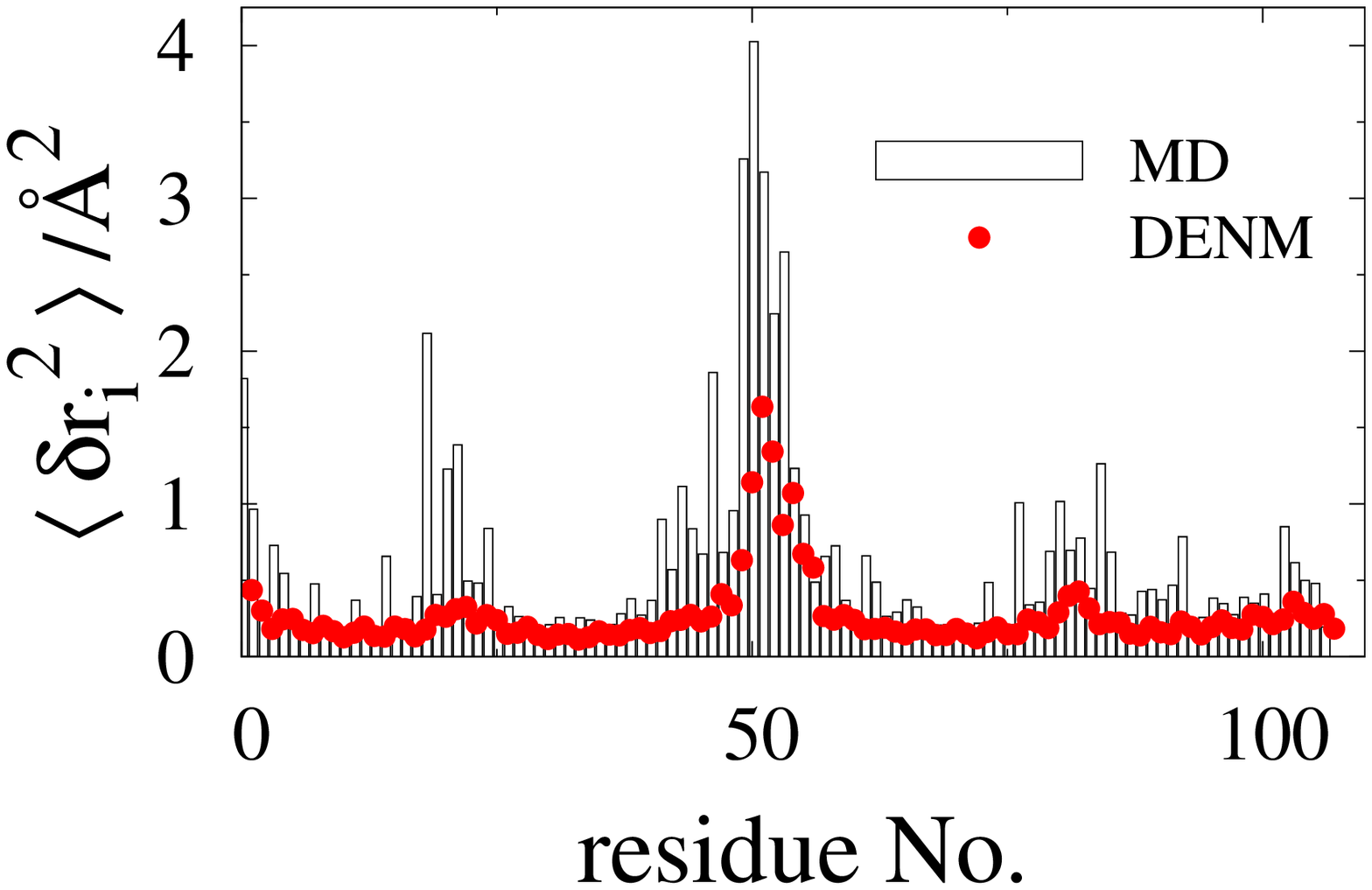}}
\hspace{0.3cm}
   \subfloat{\includegraphics*[width=5cm]{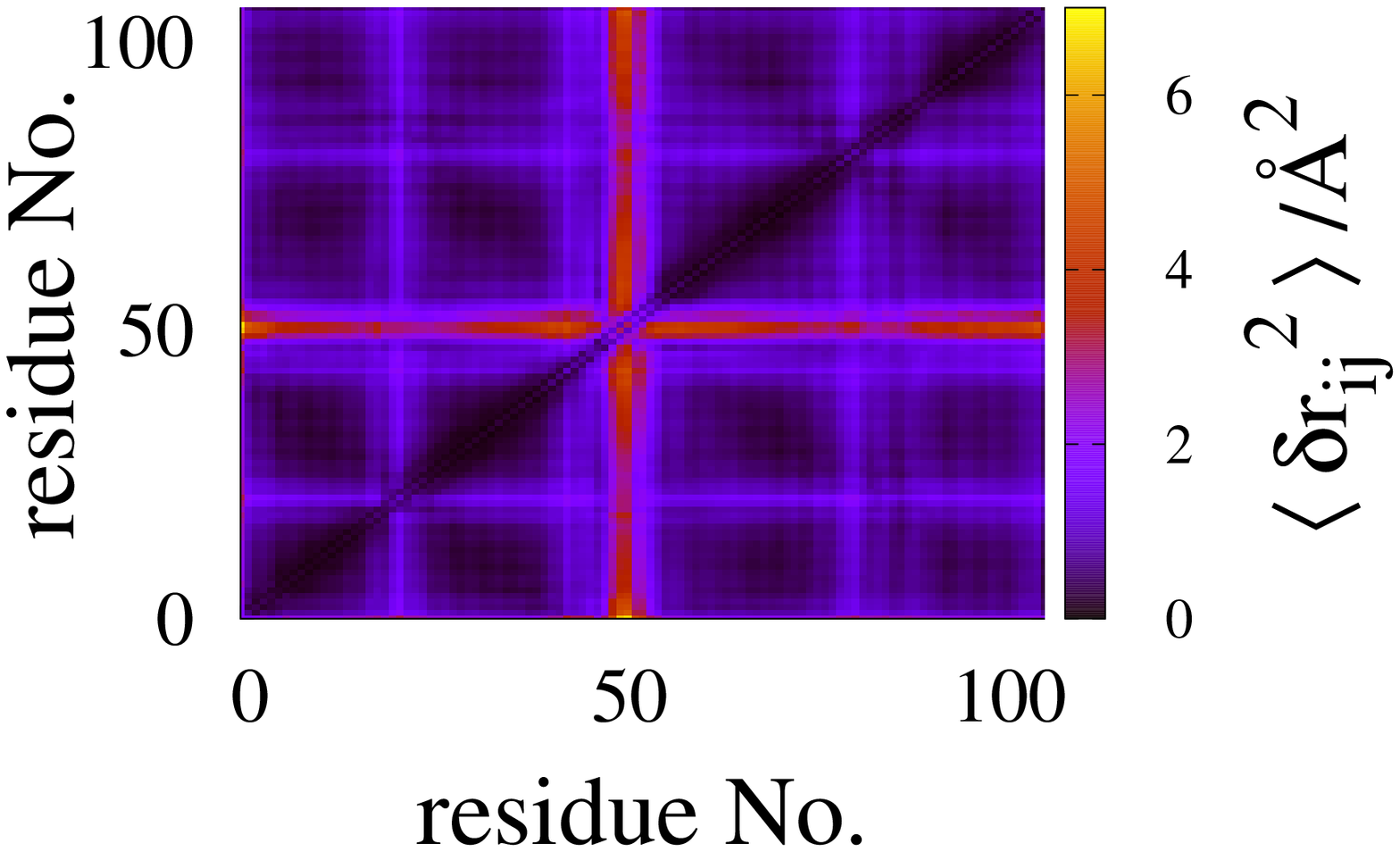}}
\hspace{0.3cm} 
\subfloat{\includegraphics*[width=5cm]{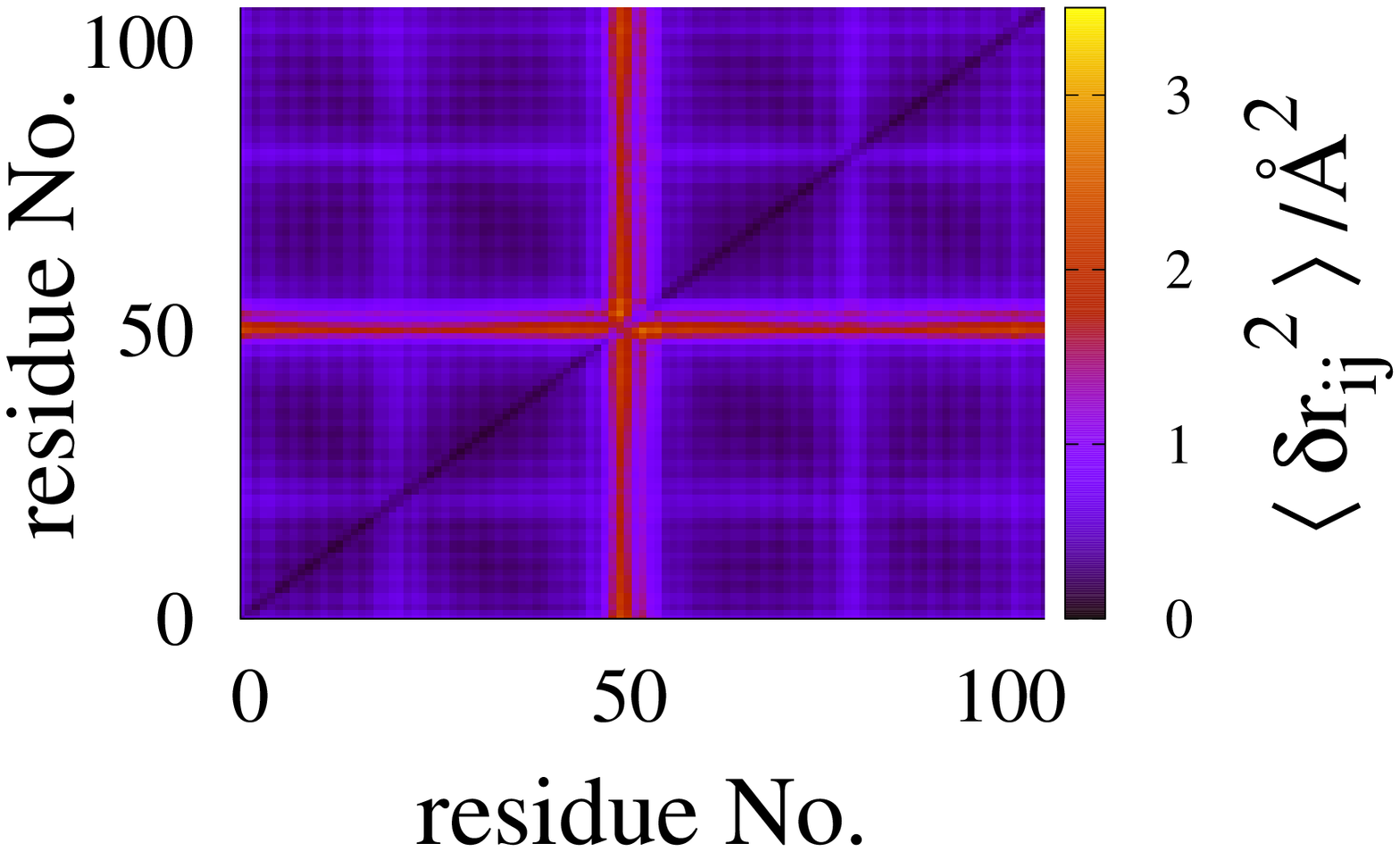}}
\caption{\small Residue mean-square displacements (msd's) $\langle (\delta
  r_i)^2\rangle$ (a) and variance-covariance matrix $\langle \delta
  \mathbf{r}_i \cdot \delta \mathbf{r}_j \rangle $ of residues'
  $C^{\alpha}$ carbons calculated from MD (b) and from DENM (c). The
  diagonals in (b) and (c) represents the residue msd's $\langle
  (\delta r_i)^2\rangle $.  The network force constant in the DENM
  calculations is $k_{\text{B}}T/ C=1$ \AA$^2$. The DENM values
  indicated by red squares in (a) need to be multiplied by about a
  factor of two in order to obtain the best-fit agreement with the MD
  data. They are separated for a better visibility in the plot. }
  \label{fig:2}
\end{figure}

\section*{RESULTS AND DISCUSSION }
\normalsize

\subsection*{Statistics and dynamics of residue displacements}

Consistent with many previous studies, we have found that elastic
networks are capable of reproducing the basic pattern of the
distribution of mean-square displacements (msd's) along the protein
backbone.  The left panel in Fig.\ \ref{fig:2} compares residue msd's
from MD with DENM calculations, while two other panels show the maps
of variance-covariance matrices. Overall, there is a good agreement
between the alteration in residue displacements along the backbone,
and corresponding cross-correlations, calculated from the two sets of
data. However, the network force constant required for the best fit of
the msd from MD simulations, $\simeq 0.3$ kcal/(mol \AA$^2$), is
somewhat lower than the value of $\simeq 1$ kcal/(mol \AA$^2$)
typically adopted from fitting the B-factors from crystallography
\citep{Tirion:96,Atilgan:01,Tama:2001tg}. This value is also about a
factor of two lower than $0.6$ kcal/(mol \AA$^2$) adopted below to
globally fit the variances of the field and electrostatic potential
fluctuations from MD data for all three proteins studied here.

The fitting of the force constant to the absolute msd's from MD makes
the network too soft. This outcome is expected since the elastic
network cuts high-frequency vibrations from the density of states and
thus underestimates the absolute magnitudes of the residue
displacements \cite{Bahar:10aa}. It appears more consistent to use for
the purpose of fitting only the portion of the msd related to
protein's global motions. This component can in fact be separated from
the vibrational part in the temperature dependence of the msd since
global motions of the protein become observable at high temperatures,
above the temperature of the protein dynamical transition
\cite{Gabel:02}.  This high-temperature portion of the protein msd is
about a half of the total magnitude \cite{Zaccai:00}, which is close
to the factor required to reconcile the force constants from
electrostatic variances and msd's.

\begin{figure}
  \centering
  \includegraphics*[width=9cm]{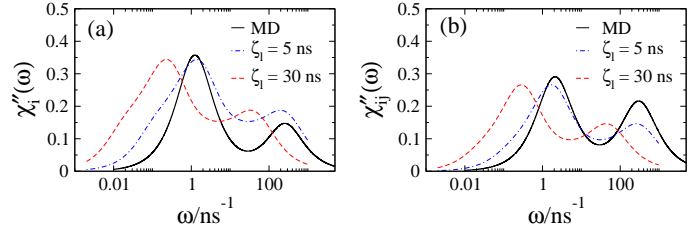}
  \caption{\small Loss spectra $\chi_{i}''(\omega)/ \chi_i'(0)$ of individual
    residues ($i=21$, a) and of pairs of residues
    $\chi_{ij}''(\omega)/ \chi_{ij}'(0)$ ($i=21$ and $j=84$, b).  The
    results have been obtained from MD (solid lines) and DENM (dashed
    lines). The calculations are done for cytB with the residues used
    in the calculations shown in Fig.\ \ref{fig:4}. The DENM
    calculations are done with $\zeta_l=30$ ns used for the
    electrostatic calculations in Fig.\ \ref{fig:5} (red dashed line)
    and with $\zeta_l=5$ ns (blue dash-dotted line). The rest of DENM
    parameters are: $\zeta_h=0.006\zeta_l$, $a=0.35$,
    $\varepsilon=100$.  }
  \label{fig:3}
\end{figure}

Figure \ref{fig:3} shows the dynamics of positions $\mathbf{r}_i(t)$
of individual residues within the network and the dynamics of
distances between residues $\mathbf{r}_{ij}(t)=\mathbf{r}_i(t) -
\mathbf{r}_j(t)$.  It reports the loss functions $\chi_i''(\omega)$
and $\chi_{ij}''(\omega)$ calculated from the self-correlation
functions of individual residues
\begin{equation}
  \label{eq:9}
  C_i(t) = \langle \delta \mathbf{r}_i(t) \cdot \delta \mathbf{r}_i(0) \rangle 
\end{equation}
and self-correlation functions of distances between the residues
\begin{equation}
  \label{eq:10}
  C_{ij}(t) =\langle \delta \mathbf{r}_{ij}(t) \cdot \mathbf{r}_{ij}(0) \rangle . 
\end{equation}
These latter correlation functions are experimentally available from
FRET \cite{Weiss:2000nx} recording the evolution of the distance between two residues
tagged with chromophores.

\begin{figure}
  \centering
  \includegraphics*[width=5cm]{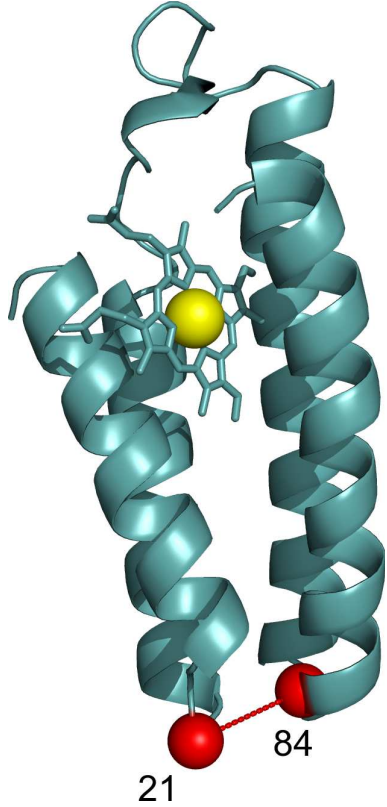}
  \caption{\small Cartoon of cytB showing the residues (red) used to produce
    the correlation functions $C_i(t)$ ($i=21$) and $C_{ij}(t)$
    ($i,j=21,84$, Eqs.\ \eqref{eq:9} and \eqref{eq:10}) and
    corresponding loss spectra $\chi_i''(\omega)$ and
    $\chi_{ij}''(\omega)$ shown in Fig.\ \ref{fig:3}. The residues
    used in producing $\chi_{ij}''(\omega)$ are connected by the red
    solid line. The red spheres representing the two chosen residues
    are centered at their corresponding $C^{\alpha}$ carbons. The Fe
    atom of the heme is rendered as a yellow sphere. }
  \label{fig:4}
\end{figure}

The comparison of $\chi_i''(\omega)$ from DENM and MD is illustrated
in Fig.\ \ref{fig:3}a for only one residue, $i=21$, belonging to cytB
protein. The pair of residues with $i=21$ and $j=84$ is used to
calculate $C_{ij}(t)$ in Fig.\ \ref{fig:3}b. The position of the pair
in the backbone of cytB is shown in Fig.\ \ref{fig:4}. Examples of
calculations for several additional pairs of residues from cytB can be
found in Supporting Materials.

Both correlation functions, $C_i(t)$ and $C_{ij}(t)$, report the
dynamics on roughly two time-scales, seen as two peaks in the their
loss functions in Fig.\ \ref{fig:3}. The two peaks represent, with
varying weights, the faster local dynamics of individual residues and
the slower global dynamics of the network.

The elastic network with single-exponential, Debye dynamics does not
capture two time-scales of the dynamics from MD simulations.  When the
Debye memory kernel $\tilde \zeta(\omega)=\zeta_0$ is used in the
response function $\chi_m(\omega)$ in Eqs.\ \eqref{eq:5} and
\eqref{eq:6}, the loss function shows only one relaxation peak,
slightly deformed from a simple Debye form by the distribution of
network's eigenvalues $\lambda_m$.  A single friction coefficient
evidently misses the fact that energy dissipation decreases with
increasing frequency of vibrations. Since the protein network is
characterized by two spring constants, lower for non-covalent
neighbors within the cutoff and higher for covalent neighbors, it must
possess at least two characteristic friction coefficients. 

Friction kernels typically used in applications are phenomenological
\cite{Hansen:03} and we as well cannot offer a consistent theoretical
formalism capturing the complex dynamics of residue displacements. We
have therefore adopted a phenomenological response function best
describing the MD results.  Instead of searching for a functional form
of $\tilde \zeta(\omega)$ reproducing MD simulations, we have resorted
to a two-Debye form for the entire response function of bead
displacements in Eq.\ \eqref{eq:5}
\begin{equation}
  \label{eq:24}
  \chi_m(\omega) = \frac{a}{i\omega\zeta_h +\lambda_m} +
  \frac{1-a}{i\omega\zeta_l +\lambda_m} .
\end{equation}
Here, the amplitude $a$ represents the relative weight of the
high-frequency dissipation with the high-frequency friction
$\zeta_h$. Correspondingly, $\zeta_l$ is the low-frequency friction,
which is larger in magnitude. It is obvious that Eq.\ \eqref{eq:24}
satisfies the static limit $\chi_m(0)=\lambda_m^{-1}$.

The two-Debye form of $\chi_m(\omega)$ is directly applied to calculate the
response function of residue displacements 
\begin{equation}
   \label{eq:23}
   \chi_{ij}(\omega) = \chi_{ij}^{\alpha\alpha}(\omega),    
\end{equation}
where the function $\chi_{ij}^{\alpha\beta}(\omega)$ is given by Eq.\
\eqref{eq:6}.  The loss spectra of residue displacements show two
peaks and qualitatively agree with the simulations. The positions and
relative heights of the peaks can be adjusted by choosing
$\zeta_{h,l}$ and the amplitude $a$ in Eq.\ \eqref{eq:24}, as is
illustrated in Fig.\ \ref{fig:3}a by dashed and dash-dotted lines. The
model can therefore be well parameterized to reproduce the dynamics of
displacements of a small subset of residues.  However, it fails to
discriminate between the dynamics of residues across the protein
backbone.

The loss functions calculated from MD for a number of residues (see
Supporting Materials) display similar two-peaks pattern, but the
weights and positions of their peaks vary among the residues. In
contrast, the dynamics of displacements of individual residues in the
elastic network are mostly driven by the global motions of the entire
network. As a result, there is little difference between the loss
functions of individual residues calculated with DENM. The same
statement applies to the dynamics of distances between residues shown
in Fig.\ \ref{fig:3}b. While one can reproduce the loss function of a
chosen pair of residues by adjusting the parameters of
$\chi_m(\omega)$ in Eq.\ \eqref{eq:24}, these parameters do not
translate to all pairs in the network and instead can be only viewed
as an average representation of the pairs dynamics.

The fast component of the dynamics, prominent in the loss function
$\chi_X''(\omega)$, is less significant in the function
$\alpha_X(\omega)$ due to the $1/\omega$ scaling of $\chi_X''(\omega)$
in Eq.\ \eqref{eq:20}.  The function $\alpha_X(\omega)$ is more
relevant to problems related to the generalized compliance as defined
by Eq.\ \eqref{eq:19}. The correct representation of the fast dynamics
is therefore of lesser importance for these type of problems.  The $1/
\omega$ scaling difference between $\chi_X''(\omega)$ and
$\alpha_X(\omega)$ explains the relative success of the network models
in reproducing the pattern of residue msd's (Fig.\ \ref{fig:2}). Those
are obtained by frequency integration of $\alpha_r(\omega)$ and are
less sensitive to the details of local dynamics of individual beads in
the network represented by high-frequency peak of their loss
functions.

\begin{figure}
\includegraphics*[width=9cm]{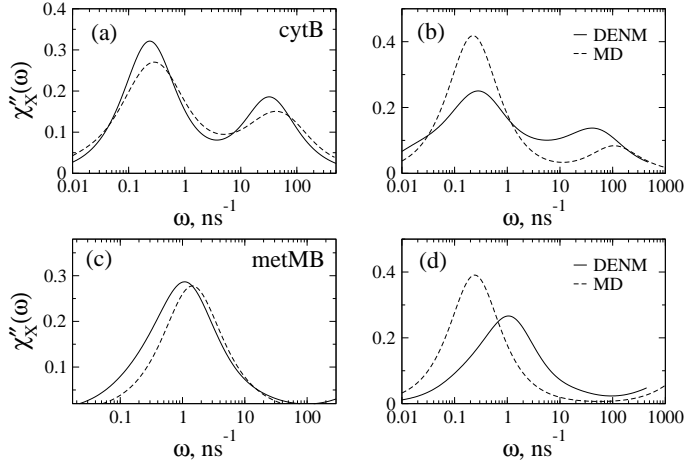}
\caption{\small Loss spectra $\chi_{\phi}''(\omega)/\chi'_{\phi}(0)$ (a,c)
  and $\chi_E''(\omega)/ \chi_E'(0)$ (b,d) for cytB (a,b) and metMB
  (c,d). The results are from MD trajectories (dashed lines) and from
  the DENM calculations (solid lines).  The DENM calculations were
  done with the two-Debye $\chi_m(\omega)$ in Eq.\ \eqref{eq:24}. The
  two-Debye relaxation parameters are: 
  $\zeta_l=30$ ns, $\zeta_h=0.006\zeta_l$, $a=0.35$ for cytB and 
  $\zeta_l=10$ ns, $\zeta_h=0.0002\zeta_l$, $a=0.35$ for metMB. The
  elastic network is defined with $k_{\text{B}}T/C = 1$ \AA$^2$,
  $\varepsilon=100$, and the cutoff radius of 15 \AA. }
\label{fig:5}
\end{figure}

\subsection*{Dynamics of the electrostatic field and potential}
Essentially all time correlation functions obtained here from MD
simulations show a pattern that is roughly represented by a
two-exponential decay, requiring in some cases a third decaying
exponent for a better mathematical fit. The relative weights of the
fast and slow components of the correlation functions differ depending
on the observable.  The loss function of the electric field
$\chi_E''(\omega)$ at the position of protein's heme is mostly
single-exponential, with the low-frequency peak much exceeding the
high-frequency one. This pattern is less uniform for the loss function
of the electrostatic potential showing different weights of the low-
and high-frequency peaks in $\chi_{\phi}''(\omega)$ among the proteins
\cite{DMjpcb:11}. Figure \ref{fig:5} shows the loss spectra
$\chi_X''(\omega) / \chi_X'(0)$ calculated from the DENM and MD for
cytB and metMB proteins. Similarly to the case with the residue
displacements, the low-frequency coefficient needs adjustment to
reproduce the position of the main loss peak: the value of
$\zeta_l=30$ ns was adopted for cytB and $\zeta_l=10$ ns was taken for
metMB. Still, the set of parameters taken to reproduce the loss
spectrum of electrostatic potential does not perform as well when
applied to the electric field (cf.\ Figs.\ \ref{fig:5}c and
\ref{fig:5}d).

\begin{figure}
  \includegraphics*[width=9cm]{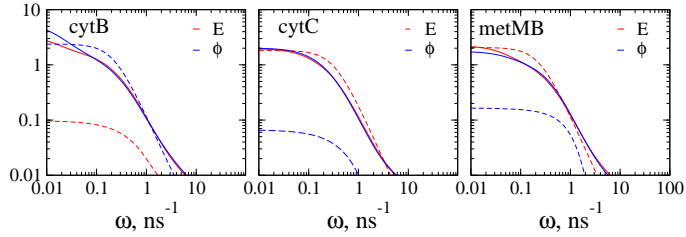}
  \caption{\small Functions $\alpha_X(\omega)$ (ns) for cytB, cytC, and metMB
    from MD (dashed lines) and DENM (solid lines). The results for the
    electrostatic potential ($\alpha_{\phi}(\omega)$, blue) and
    electric filed ($\alpha_E(\omega)$, red) are shown. A single set
    of network parameters is used for all three heme proteins:
    $\zeta_l=30$ ns, $\zeta_h=0.006\zeta_l$, $a=0.35$,
    $\varepsilon=100$. }
  \label{fig:6}
\end{figure}

The relative contribution of the fast dynamics component is reduced in
the spectral function $\alpha_X(\omega)$ describing the frequency
variation of the generalized compliance (Eqs.\ (\ref{eq:19} and
(\ref{eq:20})). Figure \ref{fig:6} compares $\alpha_X(\omega)$ from
DENM with MD data for all three globular proteins studied here. In
order to show the ability of DENM to describe the entire set of MD
data, a single set of network and friction parameters has been
assigned to all three proteins. As expected, the performance of DENM
is better for $\alpha_X(\omega)$. The average relaxation time $\langle
\tau \rangle$, given by the $\omega\rightarrow 0$ limit of
$\alpha_X(\omega)$ (Eq.\ \eqref{eq:21}), is well reproduced by the
network calculations. The deviations for some observables correspond
to the cases when the dynamics are strongly dominated by its
high-frequency component. For these cases, the fit can be improved by
adjusting the weight $a$ of the fast component in Eq.\ \eqref{eq:24}
to a higher value.

The loss functions $\chi_X(\omega)$ shown in Figs.\ \ref{fig:3},
\ref{fig:5}, and \ref{fig:6} are normalized with the corresponding
values of $\chi_X'(0)$ and therefore are not affected by the magnitude
of the spring constant $C$ describing the non-covalent interactions
within the cutoff distance $r_c=15$ \AA. The absolute values of the
variances are, however, proportional to $k_{\text{B}}T/C$, as for
instance in Eq.\ \eqref{eq:1}. We adopted in our calculations the
value of $k_{\text{B}}T/C= 1$ \AA$^2$, which is equivalent to $C\simeq
0.6$ kcal/(mol \AA$^2$). The results of calculating the generalized
compliances $\lambda_X$ for electrostatic observables (Eqs.\
\eqref{eq:19} and \eqref{eq:14}) with this choice of the force
constant are summarized in Table \ref{tab:1}. The agreement is
semi-quantitative at best. While the value of the force constant can
clearly be adjusted in each particular case, this change does not
propagate into equally good agreement between $\lambda_{\phi}$ and
$\lambda_{F}$, where $F=eE$ is the electrostatic force acting on the
unit charge placed at Fe of the heme. The requirement to reproduce
electrostatic forces might be more stringent than to describe the
electrostatic potential because of a shorter range and higher
sensitivity to the local structure of the former. Since the
statistical variances listed in Table \ref{tab:1} are not affected by
the modeling of dissipative dynamics, the properties of the network
itself need to be further fine-tuned. The inclusion of softening of
the residue displacements by electrostatic water fluctuations
\cite{DMpre:11} appears to be the first avenue for the model
improvement. This addition will allow one to accommodate for the
effect of the interface, in addition to the motions dictated by the
shape and mass distribution which elastic models capture in the first
place \citep{Halle:02,Lu:2005mi,Tama:06}.  Incorporating the water
response will effectively produce a more heterogeneous distribution of
the network force constants.

\begin{table}
  \centering
  \caption{Generalized compliances (Eq.\ \eqref{eq:19}) for the electrostatic
    potential fluctuations $\lambda_{\phi}$ (also known as reorganization
    energies of redox half reaction $\lambda=e^2\lambda_{\phi}$, kcal/mol) 
    and for the electric force, $F=eE$
    ($\lambda_F$, kcal/(mol \AA$^2$)) for the three proteins used in 
    the case study; the network is assigned
    the force constant of $C=0.6$ kcal/(mol \AA$^2$) for non-covalent springs and $\varepsilon C$, $\varepsilon=100$ 
    for covalently bound neighbors. The cutoff radius is 15 \AA.   }
  \label{tab:1}
  \begin{tabular}{lcccc}
\hline
Protein          &  $e^2\lambda_{\phi}$ (DENM) & $e^2\lambda_{\phi}$ (MD) & $\lambda_F$ (DENM) & $\lambda_F$ (MD) \\
\hline
MetMB      &   85    &  228  & 436   & 243 \\   
CytB       &   115   &  136  &   9 &  17 \\
CytC       &   141   &  159  &  18 &   83 \\
\hline    
  \end{tabular}
\end{table}

\section*{CONCLUSIONS}

Network models of biomolecules are coarse-grained representations
designed to describe their large-scale collective conformational
motions \cite{Bahar:10aa}. They capture the basic topology and packing
of residues in a biopolymer and robustly reproduce conformational
global motions driven by the distribution of mass and shape
\citep{Halle:02,Lu:2005mi,Tama:06}.  Electrostatic properties is yet
another area where coarse-graining of biomolecules might be
efficient. The long range of Coulomb interactions effectively averages
out details of the local structure suggesting that one can potentially
describe the statistics and long-time dynamics of electrostatic
fluctuations by global motions of charges assigned to the elastic
network. The demonstration that this approach is in principle
consistent with all-atom MD simulations is the main result of this
study. Two components are critical to our approach: force-field atomic
charges distributed at the residues of the network and
two-exponential, overdamped dynamics assigned to each normal mode
diagonalizing the network Hessian.

Harmonic mechanical motions of the network of beads clearly do not
incorporate dissipative dynamics of biomolecules in solution
\cite{ansari:1774,Erkip:04,Miller:2008uq} and much is still needed to
be done to achieve a physically consistent description of the protein
dynamics. The elastic network employed here assigns weaker springs
between all non-covalent neighbors within a cutoff radius and stronger
springs between covalent neighbors. Correspondingly, two global
friction coefficients are assigned to each normal mode diagonalizing
the network Hessian. This phenomenological model qualitatively
captures the two-peak loss spectrum of residue displacements and
qualitatively similar loss spectra of the electrostatic potential and
electric field.  However, the characteristic adjustable friction
coefficients used for the electrostatic fluctuations are not
transferrable to the network displacements. A new set of parameters is
needed when each property is considered separately.

\vskip 1cm
{\small This research was supported by the National Science
Foundation (DVM, CHE-0910905). CPU time was provided by the National
Science Foundation through TeraGrid resources (TG-MCB080116N).}

\bibliography{chem_abbr,dielectric,dm,statmech,protein,liquids,solvation,dynamics,glass,elastic,simulations,surface,bioet,et,nano,photosynthNew,enm,bioenergy,nih}

\begin{thebibliography}{67}
\providecommand{\url}[1]{\texttt{#1}}
\providecommand{\urlprefix}{ }

\bibitem[Fenimore et~al.(2004)Fenimore, Frauenfelder, McMahon, and
  Young]{Fenimore:04}
Fenimore, P.~W., H.~Frauenfelder, B.~H. McMahon, and R.~D. Young, 2004.
\newblock Bulk-solvent and hydration-shell fluctuations, similar to $\alpha$
  and $\beta$-fluctuations in glasses, control protein motion and functions.
\newblock \emph{Proc. Natl. Acad. Sci.} 101:14408--14413.

\bibitem[Henzlel-Wildman and Kern(2007)]{HenzlelWildman:07}
Henzlel-Wildman, K., and D.~Kern, 2007.
\newblock Dynamic personalities of proteins.
\newblock \emph{Nature} 450:964.

\bibitem[Frauenfelder et~al.(2009)Frauenfelder, Chen, Berendzen, Fenimore,
  Jansson, McMahon, Stroe, Swenson, and Young]{Frauenfelder:09}
Frauenfelder, H., G.~Chen, J.~Berendzen, P.~W. Fenimore, H.~Jansson, B.~H.
  McMahon, I.~R. Stroe, J.~Swenson, and R.~D. Young, 2009.
\newblock A unified model of protein dynamics.
\newblock \emph{Proc.\ Natl.\ Acad.\ Sci.} 106:5129--5134.

\bibitem[Marlow et~al.(2010)Marlow, Dogan, Frederick, Valentine, and
  Wand]{Marlow:2010bh}
Marlow, M.~S., J.~Dogan, K.~K. Frederick, K.~G. Valentine, and A.~J. Wand,
  2010.
\newblock The role of conformational entropy in molecular recognition by
  calmodulin.
\newblock \emph{Nat Chem Biol} 6:352--358.
\newblock \urlprefix\url{http://dx.doi.org/10.1038/nchembio.347}.

\bibitem[Warshel et~al.(2006)Warshel, Sharma, Kato, and Parson]{Warshel:06}
Warshel, A., P.~K. Sharma, M.~Kato, and W.~W. Parson, 2006.
\newblock Modeling electrostatic effects in proteins.
\newblock \emph{Biochim.\ Biophys.\ Acta} 1764:1647--1676.

\bibitem[Marcus and Sutin(1985)]{MarcusSutin}
Marcus, R.~A., and N.~Sutin, 1985.
\newblock Electron transfer in chemistry and biology.
\newblock \emph{Biochim. Biophys. Acta} 811:265--322.

\bibitem[LeBard and Matyushov(2010{\natexlab{a}})]{DMpccp:10}
LeBard, D.~N., and D.~V. Matyushov, 2010.
\newblock Protein-water electrostatics and principles of bioenergetics.
\newblock \emph{Phys.\ Chem.\ Chem.\ Phys.} 12:15335.

\bibitem[Khodadadi et~al.(2010)Khodadadi, Roh, Kisliuk, Mamontov, Tyagi,
  Woodson, Briber, and Sokolov]{Khodadadi:2010qf}
Khodadadi, S., J.~H. Roh, A.~Kisliuk, E.~Mamontov, M.~Tyagi, S.~A. Woodson,
  R.~M. Briber, and A.~P. Sokolov, 2010.
\newblock Dynamics of Biological Macromolecules: Not a Simple Slaving by
  Hydration Water.
\newblock \emph{Biophys.\ J.} 98:1321--1326.

\bibitem[Ebbinghaus et~al.(2007)Ebbinghaus, Kim, Heyden, Yu, Heugen, Gruebele,
  Leitner, and Havenith]{Ebbinghaus:07}
Ebbinghaus, S., S.~J. Kim, M.~Heyden, X.~Yu, U.~Heugen, M.~Gruebele, D.~M.
  Leitner, and M.~Havenith, 2007.
\newblock An extended dynamical hydration shell around proteins.
\newblock \emph{Proc.\ Natl.\ Acad.\ Sci.} 104:20749--20752.

\bibitem[Perticaroli et~al.(2010)Perticaroli, Comez, Paolantoni, Sassi, Lupi,
  Fioretto, Paciaroni, and Morresi]{Perticaroli:10}
Perticaroli, S., L.~Comez, M.~Paolantoni, P.~Sassi, L.~Lupi, D.~Fioretto,
  A.~Paciaroni, and A.~Morresi, 2010.
\newblock Broadband depolarized light scattering study of diluted protein
  aqueous solutions.
\newblock \emph{J.\ Phys.\ Chem. B} 114:8262.

\bibitem[Richardson and Richardson(2002)]{Richardson:02}
Richardson, J.~S., and D.~C. Richardson, 2002.
\newblock Natural $\beta$-sheet proteins use negative design to avoid
  edge-to-edge aggregation.
\newblock \emph{Proc.\ Natl.\ Acad.\ Sci.} 99:2754.

\bibitem[Lawrence et~al.(2007)Lawrence, Phillips, and Liu]{Lawrence:07}
Lawrence, M.~S., K.~J. Phillips, and D.~R. Liu, 2007.
\newblock Supercharging Proteins Can Impart Unusual Resilience.
\newblock \emph{J. Am. Chem. Soc.} 129:10110.

\bibitem[Pace et~al.(2009)Pace, Grimsley, and Scholtz]{Pace:09}
Pace, C.~N., G.~R. Grimsley, and J.~M. Scholtz, 2009.
\newblock Protein ionizable groups: pK values and their contribution to protein
  stability and solubility.
\newblock \emph{J.\ Biol.\ Chem.} 284:13285.

\bibitem[Matyushov and Morozov(2011)]{DMpre:11}
Matyushov, D.~V., and A.~Y. Morozov, 2011.
\newblock Electrostatic fluctuations promote dynamical transition in proteins.
\newblock \emph{Phys. Rev. E} 84:011908.

\bibitem[Andreatta et~al.(2005)Andreatta, P{{\'e}}rez, Kovalenko, Ernsting,
  Murphy, Coleman, and Berg]{Berg:05}
Andreatta, D., J.~L. P{{\'e}}rez, S.~A. Kovalenko, N.~P. Ernsting, C.~J.
  Murphy, R.~S. Coleman, and M.~A. Berg, 2005.
\newblock Power-law solvation dynamics in {DNA} over six decades in time.
\newblock \emph{J. Am. Chem. Soc.} 127:7270.

\bibitem[Tripathy and Beck(2010)]{Tripathy:10}
Tripathy, J., and W.~F. Beck, 2010.
\newblock Nanosecond-Regime Correlation Time Scales for Equilibrium Protein
  Structural Fluctuations of Metal-Free Cytochrome c from Picosecond
  Time-Resolved Fluorescence Spectroscopy and the Dynamic Stokes Shift.
\newblock \emph{J.\ Phys.\ Chem. B} 114:15958--15968.

\bibitem[Zhang et~al.(2007)Zhang, Wang, Kao, Qiu, Yang, Okobiah, and
  Zhong]{Zhang:07}
Zhang, L., L.~Wang, Y.-T. Kao, W.~Qiu, Y.~Yang, O.~Okobiah, and D.~Zhong, 2007.
\newblock Mapping hydration dynamics around a protein surface.
\newblock \emph{Proc.\ Natl.\ Acad.\ Sci.} 104:18461.

\bibitem[Tirion(1996)]{Tirion:96}
Tirion, M.~M., 1996.
\newblock Large amplitude elastic motions from single-parameter, atomic
  analysis.
\newblock \emph{Phys. Rev. Lett.} 77:1905--1908.

\bibitem[Atilgan et~al.(2001)Atilgan, Durell, Jernigan, Demirel, Keskin, and
  Bahar]{Atilgan:01}
Atilgan, A.~R., S.~R. Durell, R.~L. Jernigan, M.~C. Demirel, O.~Keskin, and
  I.~Bahar, 2001.
\newblock Anisotropy of Fluctuation Dynamics of Proteins with an Elastic
  Network Model.
\newblock \emph{Biophys. J.} 80:505--515.

\bibitem[Tama and Sanejouand(2001)]{Tama:2001tg}
Tama, F., and Y.~H. Sanejouand, 2001.
\newblock Conformational change of proteins arising from normal mode
  calculations.
\newblock \emph{Protein Eng} 14:1--6.

\bibitem[Tama et~al.(2003)Tama, Valle, Frank, and Brooks]{Tama:2003hc}
Tama, F., M.~Valle, J.~Frank, and r.~Brooks, C.~L., 2003.
\newblock Dynamic reorganization of the functionally active ribosome explored
  by normal mode analysis and cryo-electron microscopy.
\newblock \emph{Proc.\ Natl.\ Acad.\ Sci.\ USA} 100:9319--23.

\bibitem[Tama and Brooks(2005)]{Tama:2005oq}
Tama, F., and C.~L. Brooks, 2005.
\newblock Diversity and identity of mechanical properties of icosahedral viral
  capsids studied with elastic network normal mode analysis.
\newblock \emph{J. Mol. Biol.} 345:299--314.

\bibitem[Lu et~al.(2006)Lu, Poon, and Ma]{Lu:2006cr}
Lu, M., B.~Poon, and J.~Ma, 2006.
\newblock A New Method for Coarse-Grained Elastic Normal-Mode Analysis.
\newblock \emph{J. Chem. Theory Comput.} 2:464--471.

\bibitem[Moritsugu and Smith(2007)]{Moritsugu:2007ff}
Moritsugu, K., and J.~C. Smith, 2007.
\newblock Coarse-Grained Biomolecular Simulation with REACH: Realistic
  Extension Algorithm via Covariance Hessian.
\newblock \emph{Biophys.\ J.} 93:3460--3469.

\bibitem[Lu and Ma(2008)]{Lu:2008dq}
Lu, M., and J.~Ma, 2008.
\newblock A minimalist network model for coarse-grained normal mode analysis
  and its application to biomolecular x-ray crystallography.
\newblock \emph{Proc. Nat. Acad. Sci.} 105:15358--63.

\bibitem[Riccardi et~al.(2009)Riccardi, Cui, and Phillips]{Riccardi:09}
Riccardi, D., Q.~Cui, and G.~N. Phillips, 2009.
\newblock Application of elastic network models to proteins in the crystalline
  state.
\newblock \emph{Biophys. J.} 96:464--475.

\bibitem[Bahar et~al.(2010)Bahar, Lezon, Yang, and Eyal]{Bahar:10aa}
Bahar, I., T.~R. Lezon, L.-W. Yang, and E.~Eyal, 2010.
\newblock Global Dynamics of Proteins: Bridging Between Structure and Function.
\newblock \emph{Ann. Rev. Biophys.} 39:23--42.

\bibitem[Halle(2002)]{Halle:02}
Halle, B., 2002.
\newblock Flexibility and packing in proteins.
\newblock \emph{Proc.\ Natl.\ Acad.\ Sci.} 99:1274--1279.

\bibitem[Lu and Ma(2005)]{Lu:2005mi}
Lu, M., and J.~Ma, 2005.
\newblock The Role of Shape in Determining Molecular Motions.
\newblock \emph{Biophys. J.} 89:2395--2401.

\bibitem[Tama and Brooks(2006)]{Tama:06}
Tama, F., and C.~L. Brooks, 2006.
\newblock Symmetry, Form, and Shape: Guiding Principles for Robustness in
  Macromolecular Machines.
\newblock \emph{Annu. Rev. Biophys. Biomol. Struct.} 35:115--133.

\bibitem[Petrone and Pande(2006)]{Petrone:06}
Petrone, P., and V.~S. Pande, 2006.
\newblock Can conformational change be described by only a few normal modes?
\newblock \emph{Biophys. J.} 90:1583--1593.

\bibitem[Lyman et~al.(2008)Lyman, Pfaendtner, and Voth]{Lyman:08}
Lyman, E., J.~Pfaendtner, and G.~A. Voth, 2008.
\newblock Systematic multiscale parametrization of heterogeneous elastic
  network models of proteins.
\newblock \emph{Biophys. J.} 95:4183.

\bibitem[Hinsen and Kneller(1999)]{hinsen:10766}
Hinsen, K., and G.~R. Kneller, 1999.
\newblock A simplified force field for describing vibrational protein dynamics
  over the whole frequency range.
\newblock \emph{J. Chem. Phys.} 111:10766--10769.

\bibitem[Miller et~al.(2008)Miller, Zheng, Venable, Pastor, and
  Brooks]{Miller:2008uq}
Miller, B.~T., W.~Zheng, R.~M. Venable, R.~W. Pastor, and B.~R. Brooks, 2008.
\newblock Langevin Network Model of Myosin.
\newblock \emph{J. Phys. Chem. B} 112:6274--6281.

\bibitem[Matyushov(2011)]{DMjpcb:11}
Matyushov, D.~V., 2011.
\newblock Nanosecond Stokes Shift Dynamics, Dynamical Transition, and Gigantic
  Reorganization Energy of Hydrated Heme Proteins.
\newblock \emph{J. Phys. Chem. B} 115:10715--10724.

\bibitem[LeBard and Matyushov(2009)]{DMjpcb:09}
LeBard, D.~N., and D.~V. Matyushov, 2009.
\newblock Energetics of bacterial photosynthesis.
\newblock \emph{J.\ Phys.\ Chem. B} 113:12424--12437.

\bibitem[Tozzini(2010)]{Tozzini:11}
Tozzini, V., 2010.
\newblock Minimalist models for proteins: a comparative analysis.
\newblock \emph{Quarterly Reviews of Biophysics} 43:333--371.

\bibitem[Hinsen(2008)]{Hinsen:2008bh}
Hinsen, K., 2008.
\newblock Structural flexibility in proteins: impact of the crystal
  environment.
\newblock \emph{Bioinformatics} 24:521--8.

\bibitem[Gerek and Ozkan(2011)]{Gerek:2011ly}
Gerek, Z.~N., and S.~B. Ozkan, 2011.
\newblock Change in allosteric network affects binding affinities of PDZ
  Domains: Analysis through Perturbation Response Scanning.
\newblock \emph{PLoS Comput. Biol.} 7:e1002154.

\bibitem[Ming and Wall(2005)]{Ming:2005fj}
Ming, D., and M.~E. Wall, 2005.
\newblock Allostery in a Coarse-Grained Model of Protein Dynamics.
\newblock \emph{Phys. Rev. Lett.} 95:198103.

\bibitem[Gerek et~al.(2009)Gerek, Keskin, and Ozkan]{Gerek:2009ve}
Gerek, Z.~N., O.~Keskin, and S.~B. Ozkan, 2009.
\newblock Identification of specificity and promiscuity of PDZ domain
  interactions through their dynamic behavior.
\newblock \emph{Proteins} 77:796--811.

\bibitem[Smith(1991)]{Smith:91}
Smith, J.~C., 1991.
\newblock Protein dynamics: comparison of simulations with inelastic neutron
  scattering experiments.
\newblock \emph{Quat.\ Rev.\ Biophys.} 24:227.

\bibitem[Cametti et~al.(2011)Cametti, Marchetti, Gambi, and
  Onori]{Cametti:2011ys}
Cametti, C., S.~Marchetti, C.~M.~C. Gambi, and G.~Onori, 2011.
\newblock Dielectric Relaxation Spectroscopy of Lysozyme Aqueous Solutions:
  Analysis of the delta-Dispersion and the Contribution of the Hydration Water.
\newblock \emph{J. Phys. Chem. B} 115:7144--7153.

\bibitem[Lamm and Szabo(1986)]{Lamm:86}
Lamm, G., and A.~Szabo, 1986.
\newblock Langevine modes of macromolecules.
\newblock \emph{J.\ Chem.\ Phys.} 85:7334.

\bibitem[Ansari(1999)]{ansari:1774}
Ansari, A., 1999.
\newblock Langevin modes analysis of myoglobin.
\newblock \emph{J. Chem. Phys.} 110:1774--1780.

\bibitem[Erkip and Erman(2004)]{Erkip:04}
Erkip, A., and B.~Erman, 2004.
\newblock Dynamics of large-scale fluctuations in native proteins. Analysis
  based on harmonic inter-residue potentials and random external noise.
\newblock \emph{Polymer} 45:641--648.

\bibitem[Hansen and McDonald(2003)]{Hansen:03}
Hansen, J.~P., and I.~R. McDonald, 2003.
\newblock Theory of Simple Liquids.
\newblock Academic Press, Amsterdam.

\bibitem[Essiz and Coalson(2009)]{Essiz:2009fk}
Essiz, S.~G., and R.~D. Coalson, 2009.
\newblock Dynamic Linear Response Theory for Conformational Relaxation of
  Proteins.
\newblock \emph{J. Phys. Chem. B} 113:10859--10869.

\bibitem[Soheilifard et~al.(2011)Soheilifard, Makarov, and
  Rodin]{Soheilifard:2011pi}
Soheilifard, R., D.~E. Makarov, and G.~J. Rodin, 2011.
\newblock Rigorous coarse-graining for the dynamics of linear systems with
  applications to relaxation dynamics in proteins.
\newblock \emph{J. Chem. Phys.} 135.

\bibitem[Skep{\"o} et~al.(2006)Skep{\"o}, Linse, and Arnebrant]{Skepo:2006gf}
Skep{\"o}, M., P.~Linse, and T.~Arnebrant, 2006.
\newblock Coarse-Grained Modeling of Proline Rich Protein 1 (PRP-1) in Bulk
  Solution and Adsorbed to a Negatively Charged Surface.
\newblock \emph{J. Phys. Chem. B} 110:12141--12148.

\bibitem[Pizzitutti et~al.(2007)Pizzitutti, Marchi, and
  Borgis]{Pizzitutti:2007ul}
Pizzitutti, F., M.~Marchi, and D.~Borgis, 2007.
\newblock Coarse-Graining the Accessible Surface and the Electrostatics of
  Proteins for Protein-Protein Interactions.
\newblock \emph{J. Chem. Theory and Comp.} 3:1867--1876.

\bibitem[Leherte and Vercauteren(2009)]{Leherte:2009fr}
Leherte, L., and D.~P. Vercauteren, 2009.
\newblock Coarse Point Charge Models For Proteins From Smoothed Molecular
  Electrostatic Potentials.
\newblock \emph{J. Chem. Theory Comp.} 5:3279--3298.

\bibitem[Dong et~al.(2008)Dong, Olsen, and Baker]{Dong:08}
Dong, F., B.~Olsen, and N.~A. Baker, 2008.
\newblock Methods in Cell Biology, Elsevier, volume~84, chapter Computational
  methods for biomolecular electrostatics, 843.

\bibitem[Berardi et~al.(2004)Berardi, Muccioli, Orlandi, Ricci, and
  Zannoni]{Berardi:2004pd}
Berardi, R., L.~Muccioli, S.~Orlandi, M.~Ricci, and C.~Zannoni, 2004.
\newblock Mimicking electrostatic interactions with a set of effective charges:
  a genetic algorithm.
\newblock \emph{Chemical Physics Letters} 389:373--378.

\bibitem[Chaikin and Lubensky(1995)]{ChaikinLubensky}
Chaikin, P.~M., and T.~C. Lubensky, 1995.
\newblock Principles of condensed matter physics.
\newblock Cambridge University Press, Cambridge.

\bibitem[LeBard and Matyushov(2008)]{DMjcp2:08}
LeBard, D.~N., and D.~V. Matyushov, 2008.
\newblock Redox Entropy of Plastocyanin: Developing a Microscopic View of
  Mesoscopic Solvation.
\newblock \emph{J.\ Chem.\ Phys.} 128:155106.

\bibitem[MacKerell et~al.(1998)MacKerell, Bashford, Bellott, Jr., Evanseck,
  Field, Fischer, Gao, Guo, S.~Ha, McCarthy, Kuchnir, Kuczera, Lau, Mattos,
  Michnick, Ngo, Nguyen, Prodhom, III, Roux, Schlenkrich, Smith, Stote, Straub,
  Watanabe, Wirkiewicz-Kuczera, Yin, and Karplus]{charmm22}
MacKerell, A.~D., D.~Bashford, M.~Bellott, R.~L.~D. Jr., J.~D. Evanseck, M.~J.
  Field, S.~Fischer, J.~Gao, H.~Guo, D.~S.~Ha, J.~McCarthy, L.~Kuchnir,
  K.~Kuczera, F.~T.~K. Lau, C.~Mattos, S.~Michnick, T.~Ngo, D.~T. Nguyen,
  B.~Prodhom, W.~E.~R. III, B.~Roux, M.~Schlenkrich, J.~C. Smith, R.~Stote,
  J.~Straub, M.~Watanabe, J.~Wirkiewicz-Kuczera, D.~Yin, and M.~Karplus, 1998.
\newblock All-atom empirical potential for molecular modeling and dynamics
  studies of proteins.
\newblock \emph{J.\ Phys.\ Chem. B} 102:3586--3616.

\bibitem[Phillips et~al.(2005)Phillips, Braun, Wang, Gumbart, Tajkhorshid,
  Villa, Chipot, Skeel, Kale, and Schulten]{namd}
Phillips, J.~C., R.~Braun, W.~Wang, J.~Gumbart, E.~Tajkhorshid, E.~Villa,
  C.~Chipot, R.~D. Skeel, L.~Kale, and K.~Schulten, 2005.
\newblock Scalable molecular dynamics with NAMD.
\newblock \emph{J.\ Comp.\ Chem.} 26:1781--1802.

\bibitem[Jorgensen et~al.(1983)Jorgensen, Chandrasekhar, Madura, Impey, and
  Klein]{tip3p:83}
Jorgensen, W.~L., J.~Chandrasekhar, J.~D. Madura, R.~W. Impey, and M.~L. Klein,
  1983.
\newblock Comparison of simple potential functions for simulating liquid water.
\newblock \emph{J.\ Chem.\ Phys.} 79:926--935.

\bibitem[Humphrey et~al.(1996)Humphrey, Dalke, and Schulten]{VMD}
Humphrey, W., A.~Dalke, and K.~Schulten, 1996.
\newblock VMD - Visual Molecular Dynamics.
\newblock \emph{Journal of Molecular Graphics} 14:33--38.

\bibitem[LeBard and Matyushov(2010{\natexlab{b}})]{DMjpcb:10}
LeBard, D.~N., and D.~V. Matyushov, 2010.
\newblock Ferroelectric hydration shells around proteins: Electrostatics of the
  protein-water interface.
\newblock \emph{J.\ Phys.\ Chem. B} 114:9246--9258.

\bibitem[Autenrieth et~al.(2004)Autenrieth, Tajkhorshid, Baudry, and
  Luthney-Schulten]{Autenrieth:04}
Autenrieth, F., E.~Tajkhorshid, J.~Baudry, and Z.~Luthney-Schulten, 2004.
\newblock Classical force field parameters for the heme prosthetic group of
  cytochrome \textit{c}.
\newblock \emph{J. Comp. Chem.} 25:1613.

\bibitem[Meuwly et~al.(2002)Meuwly, Becker, Stote, and Karplus]{Meuwly:02}
Meuwly, M., O.~M. Becker, R.~Stote, and M.~Karplus, 2002.
\newblock NO rebinding to myoglobin: a reactive molecular dynamics study.
\newblock \emph{Biophys. Chem.} 98:183.

\bibitem[Zhang and Straub(2009)]{ZhangStraub:09}
Zhang, Y., and J.~E. Straub, 2009.
\newblock Diversity of solvent dependent energy transfer pathways in heme
  proteins.
\newblock \emph{J.\ Phys.\ Chem. B} 113:825.

\bibitem[Gabel et~al.(2002)Gabel, Bicout, Lehnert, Tehei, Weik, and
  Zaccai]{Gabel:02}
Gabel, F., D.~Bicout, U.~Lehnert, M.~Tehei, M.~Weik, and G.~Zaccai, 2002.
\newblock Protein dynamics studied by neutron scattering.
\newblock \emph{Quat.\ Rev.\ Biophys.} 35:327--367.

\bibitem[Zaccai(2000)]{Zaccai:00}
Zaccai, G., 2000.
\newblock How soft is a protein? A protein dynamics force constant measured by
  neutron scattering.
\newblock \emph{Science} 288:1604--1607.

\bibitem[Weiss(2000)]{Weiss:2000nx}
Weiss, S., 2000.
\newblock Measuring conformational dynamics of biomolecules by single molecule
  fluorescence spectroscopy.
\newblock \emph{Nat. Struct. Biol.} 7:724 -- 729.

\end{thebibliography}

\end{document}